\def\b{{\bf b}}
\def\c{{\bf c}}
\def\v{{\bf v}}
\def\A{{\bf A}}

\def\E{{\bf E}}
\def\B{{\bf B}}
\def\D{{\bf D}}
\def\L{{\rm L}}
\def\T{{\rm T}}
\def\j{{\bf j}}
\def\x{{\bf x}}
\def\p{{\bf p}}
\def\q{{\bf q}}

\def\c{{\bf c}}

\def\bzeta{\mbox{\boldmath$\zeta$}}
\def\grad{\mbox{\boldmath$\nabla$}}
\def\half{{\textstyle{1\over2}}}

\def\CA{C_{\rm A}}
\def\eps{\epsilon}

\def\tr{{\rm tr}}

\def\g{\sigma}

\def\drangle{\rangle\!\rangle}
\def\dlangle{\langle\!\langle}

\def\Bigdrangle{\Bigr\rangle\!\Bigr\rangle}
\def\Bigdlangle{\Bigl\langle\!\Bigl\langle}

\def\deltaS{\delta^{S_2}}
\def\deltaC{\delta\hat C}

\def\Po{{\hat P}_0}

\def\PL{P_{\rm L}}
\def\PT{P_{\rm T}}

\def\sig{\bar\sigma}

\def\trans{\top}

\def\undertilde #1%
    {%
    \setbox0=\hbox {$#1$}%
    \setbox1=\hbox {$\tilde {\phantom {\copy0}}$}%
    \setbox2=\vtop {\offinterlineskip\box0\kern 1pt \box1}%
    \dp2 = 2.8 pt \box2\relax%
    }%

\def\underwick#1{{\vtop{\ialign{##\crcr
   $\hfil\displaystyle{#1}\hfil$\crcr\noalign{\kern3pt\nointerlineskip}
   \kern5pt\vrule\vbox to3pt{}\hrulefill\vrule\kern5pt
   \crcr\noalign{\kern3pt}}}}}
\def\overwick#1{{\vbox{\ialign{##\crcr\noalign{\kern3pt}
   \kern5pt\vrule\vtop to3pt{}\hrulefill\vrule\kern5pt
   \crcr\noalign{\kern3pt\nointerlineskip}
   $\hfil\displaystyle{#1}\hfil$\crcr}}}}

\def\overwickl#1{{\vbox{\ialign{##\crcr\noalign{\kern3pt}
   \kern5pt\vrule\vtop to3pt{}\hrulefill
   \crcr\noalign{\kern3pt\nointerlineskip}
   $\hfil\displaystyle{#1}\hfil$\crcr}}}}
\def\overwickr#1{{\vbox{\ialign{##\crcr\noalign{\kern3pt}
   \vtop to3pt{}\hrulefill\vrule\kern5pt
   \crcr\noalign{\kern3pt\nointerlineskip}
   $\hfil\displaystyle{#1}\hfil$\crcr}}}}

\documentstyle[preprint,aps,eqsecnum,fixes,epsfig]{revtex}
\tightenlines
\begin {document}


\preprint {UVA/Arnold--99--45}

\title
{An effective theory for $\omega \ll k \ll gT$ color dynamics
in hot non-Abelian plasmas}

\author {Peter Arnold}

\address
    {%
    Department of Physics,
    University of Virginia,
    Charlottesville, VA 22901
    }%
\date {November 1999}

\maketitle
\vskip -20pt

\begin {abstract}%
{%
  A proper sequence of effective theories, corresponding to larger and larger
  distance scales, is crucial for analyzing real-time equilibrium physics
  in hot non-Abelian plasmas.  For the study of color dynamics (by which I
  mean physics involving long wavelength gauge fluctuations), an important
  stepping stone in the sequence of effective theories is to have a good
  effective theory for dynamics with wave number $k$ well below the Debye
  screening mass.  I review how such dynamics is associated with inverse
  time scales $\omega \ll k$.  I then give a compact way to package, in
  the $\omega \ll k$ limit, B\"odeker's description of $k \ll m$ physics,
  which was in terms
  of Vlasov equations with collision terms.  Finally, I show how the resulting
  effective theory can be reformulated as a path integral.
}%
\end {abstract}

\thispagestyle{empty}


\section {Introduction}

The color fluctuations of very hot, weakly-coupled, non-Abelian plasmas
are non-perturbatively large at distance scales $R$ of order $(g^2 T)^{-1}$.
Their dynamics is of particular interest because it is responsible for
the large rate of baryon number violation in hot electroweak theory, and so
lies at the heart of electroweak scenarios for baryogenesis.
``Hot'' here means hot enough to (a) be ultra-relativistic, (b) ignore
chemical potentials, and (c) be in the hot, symmetric phase if there is
a Higgs mechanism.
It is now known \cite{bodeker} that the time scale associated with
non-perturbative color dynamics is
$t \sim [g^4 T \ln(1/g)]^{-1}$, which is long in the sense that
$t \gg R$ (in the weakly-coupled limit).
Equivalently, the spatial momentum and the frequency scales associated
with non-perturbative color dynamics are
\begin {equation}
   k \sim g^2 T , \qquad \omega \sim g^4 T \ln(1/g) .
\end {equation}
This momentums scale $k$ is small compared to the Debye mass
\begin {equation}
   m \sim gT .
\end {equation}
The goal of this paper is to present an effective theory for
color dynamics on scales $\omega \ll k \ll m$,
to formulate that effective theory solely in terms of
gauge fields $A_\mu(t,\x)$,
and to write the effective theory in path integral form.

It has been known for some time \cite{braaten&pisarski}
how to write a leading-order effective
theory for color dynamics at the scale $k \sim m \sim gT$,
where leading-order means that corrections are suppressed by powers
of $g$.
The zero-temperature non-Abelian Maxwell equations are modified by
what are known as ``hard thermal loops,'' which incorporate the
effects of interactions of the soft $k \sim gT$ degrees of freedom
with hard $k \sim T$ thermal excitations in the plasma.
There is a standard way of writing this effective theory which has a simple
physical interpretation
\cite{adjXadj}.
One treats the soft fields classically, and replaces the hard excitations
by classical distribution functions $n(\x,\p,t)$ which describe the
density of hard excitation at position $\x$ with momentum $\p$.
Writing down Maxwell's equations, together with an appropriately
gauge-covariant, linearized Boltzmann equation for $n$, then produces the
leading-order effective theory.
$n$ is a density matrix in color space, and the piece of it that's
relevant to long-distance color dynamics (at leading order) is the
adjoint color piece.  It is also convenient and conventional to integrate this
adjoint piece over the magnitude $|\p|$ of momentum, replacing $n(\x,\p,t)$
by an adjoint field $W(\x,\v,t)$, where $\v \equiv \hat\p$.  The
resulting equations, if $W$ is given a convenient overall normalization, are
\cite{bodeker,W}
\begin {mathletters}
\begin {equation}
   (D_t + \v\cdot\D) W - \v\cdot\E = 0 ,
\label{eq:htlboltz}
\end {equation}
\begin {equation}
   D_\nu F^{\mu\nu} = j^\mu = m^2 \langle v^\mu W \rangle_\v ,
\label {eq:maxwell}
\end {equation}
\end {mathletters}%
where $m \sim gT$ is again the leading-order Debye mass,
$\langle \cdots \rangle_\v$ denotes angular averaging over the direction
$\v$, and $v^\mu \equiv (1,\v)$.
Formally solving the Boltzmann equation for $W$ and plugging the result
into the Maxwell equation, one obtains the hard-thermal loop equation of
motion for the soft gauge field, which is
\begin {equation}
   D_\nu F^{\mu\nu} = j^\mu =
   m^2 \langle v^\mu (D_t + \v\cdot\D)^{-1} \v\cdot\E \rangle_\v ,
\label {eq:htl}
\end {equation}
This equation contains, among other things, the physics of Debye
screening, which screens static electric fields over distances of
order $1/m$.

A qualitatively important point \cite{asy}
can be extracted from (\ref{eq:htl}):
$k \ll m$ physics is dominated by frequencies $\omega \ll k$.
For the sake of quickly reviewing this point here, focus for
simplicity on the linear terms on the right-side of (\ref{eq:htl}),
focus on their $\omega \ll k$ behavior, and let's check self-consistently
that the dominant frequency falls in the $\omega \ll k$ regime.
Focus in particular on the transverse modes of the gauge field, which are not
Debye screened for $k \ll m$.
In the $\omega \ll k$ limit, one can show that the spatial current
$\j$ given by the right-hand side of (\ref{eq:htl}) becomes, in
the transverse sector,
\begin {equation}
   \j_\T \simeq {\pi m^2\over 4k} \, \E_\T
        ~~+~~ \mbox{(higher order in \A)}.
\end {equation}
Fixing $A_0=0$ gauge, and working in Fourier space, Ampere's Law 
then becomes
\begin {equation}
   (-\omega^2 + k^2)\A_\T \simeq {\pi m^2\over 4k} \, i\omega\A_\T
        ~~+~~ \mbox{(higher order in \A)}.
\label {eq:puppy}
\end {equation}
The coefficient of $\A_\T$ on the right-hand side is simply the
$\omega \ll k$ limit of the transverse hard thermal loop self-energy \cite{pi}.
For $\omega \ll k$, (\ref{eq:puppy}) becomes
\begin {equation}
   k^2 A_\T \sim {m^2 \over k} \, i\omega \A_\T
\end {equation}
in orders of magnitude,
if interactions are ignored.  The characteristic frequency is then of order
\begin {equation}
   |\omega| \sim {k^3 \over m^2}
   \qquad
   \mbox{(ignoring interactions)} ,
\label{eq:omega0}
\end {equation}
and we can now verify that this frequency indeed satisfies the assumed
relationship $\omega \ll k$ when $k \ll m$.
For this reason, in discussing effective theories for $k \ll m$, it is
relevant and useful to also specialize to $\omega \ll k$.
Interactions modify the estimate (\ref{eq:omega0}) when
$k \ll \gamma$ \cite {bodeker,Blog1},
where $\gamma \sim g^2 T \ln(1/g)$ is the inverse mean
free time between color randomizing collisions, but the result
that the characteristic frequency scale $\omega$ is small
compared to $k$ is unaffected.

The theory (\ref{eq:htl}) represents an effective theory for
momentum scales small compared to $T$.
B\"odeker has discussed what happens if one goes further and integrates
out the physics down to some scale $\mu \ll m$.
The hard particles which, microscopically, make up the color distributions
$W$ can have color-randomizing collisions by $t$-channel gluon exchange.
Such collisions are dominated by momentum exchanges $q$ in the range
$g^2 T \lesssim q \lesssim m$.  Integrating out part of this
momentum range generates an explicit collision term in the Boltzmann
equation, replacing (\ref{eq:htlboltz}) by
\begin {mathletters}
\label{eq:effB}
\begin {equation}
   (D_t + \v\cdot\D) W - \v\cdot\E = - \deltaC \, W + \xi .
\label{eq:boltz}
\end {equation}
\begin {equation}
   D_\nu F^{\mu\nu} = j^\mu = m^2 \langle v^\mu W \rangle_\v .
\label {eq:maxwell1}
\end {equation}
\end {mathletters}%
$\deltaC$ is a linearized collision operator.
The magnitude of $\deltaC$ is logarithmically sensitive to the separation
of the scales $\mu$ and $m$, and B\"odeker has calculated $\deltaC$ at
leading-order in that logarithm to be the local (in $\x$) operator
defined by
\begin {mathletters}
\label {eq:deltaC}
\begin {equation}
   \deltaC \, W(\v)
   \equiv \langle \delta C(\v,\v') \, W(\v') \rangle_{\v'} ,
\end {equation}
\begin {equation}
   \delta C(\v,\v') \approx
    \gamma(\mu) \left[
       \deltaS(\v-\v')
       - {4\over\pi} \> {(\v\cdot\v')^2 \over \sqrt{1-(\v\cdot\v')^2}} 
    \right] ,
\label {eq:deltaCleading}
\end {equation}
\begin {equation}
   \gamma(\mu) \approx \CA \alpha T \, \ln\left(m \over \mu\right) .
\end {equation}
\end {mathletters}%
Here $\approx$ denotes equality at leading-log order,
meaning that corrections are down by $[\ln(m/\mu)]^{-1}$,
and $\deltaS$ is a $\delta$-function on the unit sphere, normalized so
that $\langle \deltaS(\v-\v') \rangle_\v = 1$.
To leading log order, $\gamma(\mu)$ is
what's known as the hard
thermal gluon damping rate if one sets $\mu \sim g^2 T$.
This represents
the inverse mean free path for color-randomizing collisions of the hard
particles that, microscopically, make up the color distribution $W$.

The collision term in the Boltzmann equation damps the system towards
equilibrium.  In order to describe the physics of thermal fluctuations
around equilibrium, one must also include a thermal noise term,
which is the $\xi$ shown in equation (\ref{eq:boltz}).
This equation is therefore an example of a Langevin equation.
B\"odeker derived the noise term, but one can also argue for it on general
principles based on the fluctuation-dissipation theorem (for instance,
along the lines of ref.\ \cite{Blog1} or \cite{manuelFD}).
B\"odeker found
Gaussian white noise with correlation
\begin {equation}
    \dlangle \xi^a(\v,\x,t) \, \xi^b(\v',\x',t') \drangle =
    {2 T\over m^2} \, \delta C(\v,\v') \, \delta^{ab}
    \> \delta^{(3)}(\x{-}\x') \> \delta(t{-}t') .
\label{eq:xi}
\end {equation}
In writing formulas later on, it will be convenient to suppress indices and
$\delta$ functions and write correlations like the above in the
short-hand notation
\begin {equation}
    \dlangle \xi \xi \drangle =
    {2 T\over m^2} \, \deltaC .
\end {equation}

The combination of eqs.\ (\ref{eq:effB}) and
(\ref{eq:xi}) make up B\"odeker's effective theory for $k \ll m$.
For B\"odeker, this version of the theory was merely a stepping stone
to deriving an even simpler and more infrared effective theory for
$k \ll \gamma$, where $W$ was eliminated.  That theory is of the form
\begin {mathletters}
\label{eq:bodeker}
\begin {equation}
   \D\times\B = \sigma \E + \bzeta ,
\end {equation}
\begin {equation}
    \dlangle \zeta^a_i(\x,t) \, \zeta^b_j(\x',t') \drangle =
    2 \sigma T\,\delta_{ij} \, \delta^{ab}
    \> \delta^{(3)}(\x{-}\x') \> \delta(t{-}t') .
\end {equation}
\end {mathletters}%
It has been used as the basis for numerical simulations to obtain
the leading-log result for the hot electroweak baryon number violation
rate \cite{mooreBlog}.

Now return to the previous $k \ll m$ effective theory (\ref{eq:effB}).
The purpose of this paper is to present a cleaner, tidier version
of this effective theory, more suitable for going beyond leading-log
order in calculations.
In particular, I shall (1) take the $\omega \ll k$ limit,
discussed earlier, (2) show how to eliminate $W$ from the result to
obtain a single Langevin equation for $\A$, somewhat analogous to
(\ref{eq:htl})
but with damping and noise, and (3) show how to rewrite this
Langevin equation as a path integral.

Part of the reason for wanting to take the $\omega \ll k$ limit is a
pragmatic one.
In field theory calculations, one tends to think of the philosophy
of effective theories in the language of the Wilsonian renormalization
group---``integrating out modes with $k \gtrsim \mu$.''  But a Wilson-style
approach is generally impractical for perturbative calculations.
In practice,
one usually {\it keeps}\/ modes with $k \gg \mu$ and
instead uses renormalization subtractions to achieve an equivalent result.
Typically, dimensional regularization is used to regularize
the ultraviolet.
In an effective theory for scales $k \ll \mu$, it doesn't matter much what
the physics is in the ultraviolet ($k \gg \mu$)---one adjusts the
parameters of the effective theory to correct for the difference between
the UV behavior of the effective theory and the UV behavior of the
real theory.  So, for instance, the bare $\deltaC$ in (\ref{eq:boltz})
should be set to the {\it difference}\/ between the collisions generated in the
real theory due to gluon exchange with $q > \mu$, and those generated
in the effective theory due to gluon exchange with $q > \mu$.
The difficulty with B\"odeker's $k \ll m$ effective theory as it stands
is that, if one doesn't simply throw away the $k \gg \mu$ modes
(which is difficult to do by hand in a gauge-invariant manner),
then the equations (\ref{eq:effB}) in fact
reproduce all of the complicated $k \sim m$ behavior of the original
hard thermal loop theory (\ref{eq:htl}):
plasmons, the Debye screening threshold, etc.
Because of this, there's {\it no} difference between the $q > \mu$
contribution to $\deltaC$ in the two theories, and one should set
the bare $\deltaC$ in (\ref{eq:boltz}) to zero, returning right back
to the original hard thermal loop description (\ref{eq:effB}).
For the leading-log calculations of B\"odeker, none of this mattered---one
could think of Wilsonian-style cut-offs at $k=\mu$, and all the associated
difficulties are sub-leading order.
To cleanly discuss effects beyond leading-log order, however, a more systematic
approach to the effective $k \ll m$ effective theory is required, and it
behooves us to reformulate the effective theory in a form where its
UV behavior is as simple as possible and has no structure for $k \gg \mu$.

One of the other goals of this paper will be to reformulate the
$k \ll m$ effective
theory as a path integral.
[As a warm-up, I will also review how to do the same
for the simpler $k \ll \gamma$ effective theory of (\ref{eq:bodeker}).]
One reason this is useful is that path integrals provide, for many people,
a more familiar starting point for calculations than do Langevin equations.
Another reason is that one can fix gauges for perturbative calculations by
the usual Faddeev-Popov procedure.
The theory (\ref{eq:bodeker}), for instance, was derived by B\"odeker
specifically in $A_0=0$ gauge.
By converting the $A_0=0$ gauge result into a path integral and
then generalizing the result to a gauge-invariant form, it will be
easy to see how to correctly account for other, non-ghost-free gauge
fixings, such as Coulomb gauge.
Such gauges can be very convenient for calculations.

The advantages of the formalism discussed in this paper are put into
use by me and Yaffe in ref.\ \cite{overview,sigma},
where we compute the next-to-leading-log
corrections to B\"odeker's far-infrared effective theory (\ref{eq:bodeker}),
and use it
to analyze next-to-leading-logarithm corrections to the color conductivity
and the hot electroweak baryon number violation rate.


\section {Preview of results}

I'll recap B\"odeker's original $k \ll m$ effective theory, now splitting
Maxwell's equations into Gauss' Law and Ampere's Law:
\begin {mathletters}
\begin {equation}
   (D_t + \v\cdot\D) W - \v\cdot\E = - \deltaC \, W + \xi ,
\label{eq:Ba}
\end {equation}
\begin {equation}
   \D\cdot\E = m^2 \langle W \rangle ,
\label {eq:Bb}
\end {equation}
\begin {equation}
   - D_t \E + \D\times\B = m^2 \langle \v W \rangle .
\label{eq:Bc}
\end {equation}
\end {mathletters}%
My result for appropriate equations in the $\omega \ll k$ limit,
discussed in section \ref{sec:W}, will be
\begin {mathletters}
\label {eq:R}
\begin {equation}
   \v\cdot\D W - \v\cdot\E = - \deltaC \, W + \xi ,
\label{eq:Ra}
\end {equation}
\begin {equation}
   0 = m^2 \langle W \rangle ,
\label {eq:Rb}
\end {equation}
\begin {equation}
   \D\times\B = m^2 \langle \v W \rangle .
\label{eq:Rc}
\end {equation}
\end {mathletters}%
In section \ref{sec:twoeq}, I discuss the form of Gauss' Law (\ref{eq:Rb})
and Ampere's Law (\ref{eq:Ra}) if the Boltzmann equation (\ref{eq:Ra}) is
used to eliminate $W$.
In section \ref{sec:combine}, I then go on to show how Gauss' Law
and Ampere's Law,
together with the noise correlation (\ref{eq:xi}), can be combined into
a simple form analogous to (\ref{eq:bodeker}),
\begin{mathletters}
\label{eq:combine}
\begin {equation}
   \D \times \B = \sig(\D) \, \E + \bzeta ,
\label{eq:Aeq}
\end {equation}
\begin {equation}
   \dlangle \bzeta \bzeta \drangle
   = 2T \, \sig(\D) ,
\label{eq:zz}
\end {equation}
\end {mathletters}%
where the operator $\sig(\D)$ will be defined later.
This is an example of a Langevin equation with ``multiplicative noise,''
which simply means that the noise amplitude (\ref{eq:zz}) depends on the
dynamical variable $\A$.  Such equations are notorious for being
ambiguous and sensitive to the details of ultraviolet regularization.
In section \ref{sec:path}, I will address these issues, and show how to
formulate the theory as a gauge-invariant path integral.
The path integral has the form
\begin {equation}
   Z = \int [{\cal D}A_0(\x,t)] [{\cal D}\A(\x,t)] \>
         \exp\left(-\int dt\> d^3 x \> L\right) ,
\end {equation}
\begin {equation}
   L = {1\over 4T}
           \Bigl[ -\sig(\D)\, \E + \D\times\B \Bigr]^{\rm T}
           \sig(\D)^{-1}
           \Bigl[ -\sig(\D)\, \E + \D\times\B \Bigr]
    + L_1[\A] .
\label {eq:L2}
\end {equation}
Very roughly speaking, the Gaussian integral in $-\sig(\D)\, \E + \D\times\B$
implements a Gaussian probability distribution for
$-\sig(\D)\,\E + \D\times\B$, and so implements (\ref{eq:combine}).
The term $L_1[\A]$ is a complicated factor related to a Jacobian and to
resolving the aforementioned ambiguities, and it will be discussed later.


\section {The \lowercase{$\omega \ll k$} limit of the $W$ equations}
\label{sec:W}

The $\omega \ll k$ limit of the Boltzmann equation (\ref{eq:Ba}) is
easy to understand:
we can ignore the $D_t W$ term compared to the $\v\cdot\D W$ term.
The resulting equation (\ref{eq:Ra})
is no longer an evolution equation for $W$; instead,
$W$ is determined solely by
the instantaneous values of $\E$ and $\xi$.  Formally,
\begin {mathletters}
\label {eq:W0}
\begin {equation}
   W = \hat G (\v\cdot\E + \xi)
\end {equation}
with
\begin {equation}
   \hat G \equiv (\v\cdot\D + \deltaC)^{-1} .
\label {eq:G}
\end {equation}
\end {mathletters}

Let's now analyze Gauss' Law (\ref{eq:Bb}) using this small $\omega$
approximation to $W$:
\begin {equation}
   \D\cdot\E \simeq m^2 \langle \hat G (\v\cdot\E + \xi) \rangle .
\label{eq:gauss2}
\end {equation}
Again, the notation $\langle \cdots \rangle$ indicates averaging over
$\v$-space, but one must carefully keep in mind that $\deltaC$ and
$\hat G$ are
operators in $\v$-space.  This notation is that of
ref.\ \cite{Blog1}, and the reader may find a thorough discussion
of it in the introduction of ref.\ \cite{sigma}.
It's now useful to split $\E$ into longitudinal and transverse pieces
$\E_\L$ and $\E_\T$ \cite{Blog2}, defined by the longitudinal and transverse
projection operators
\begin {mathletters}
\begin {eqnarray}
    {\PL}^{ij} &=& D^i D^{-2} D^j ,
\label {eq:PL}
\\
    {\PT}^{ij} &=& \delta^{ij} - {\PL}^{ij} ,
\label {eq:PT}
\end {eqnarray}
\end {mathletters}%
where $i$ and $j$ run over spatial indices
and $D^{-2}$ means $(\D\cdot\D)^{-1}$.
The order of magnitude of the left-hand side of (\ref{eq:gauss2}) is
then $O(k E_\L)$.  The right-hand side of (\ref{eq:gauss2}) has,
among other things, a term $m^2 \langle \hat G \v\cdot\E_\L \rangle$
involving $\E_\L$.
Using the projection operator (\ref{eq:PL}) and a frequently useful
trick \cite{sigma}, this term can be rewritten as
\begin {equation}
   m^2 \langle \hat G \v\cdot\E_\L \rangle
   = m^2 \langle \hat G \v\cdot\D \rangle D^{-2} \D\cdot\E
   = m^2 \langle \hat G (\v\cdot\D + \deltaC) \rangle D^{-2} \D\cdot\E
   = m^2 D^{-2} \D\cdot\E .
\label {eq:terml}
\end {equation}
The middle equality follows because $\deltaC$ has the property of
annihilating functions that do not depend of $\v$.
(See refs.\ \cite{bodeker,Blog2,sigma} for discussions of this.)
Therefore, as a lexical rule,
\begin {equation}
   \deltaC \rangle = {\rangle}
\end {equation}
and, similarly,
\begin {equation}
   \langle \deltaC = \langle \, .
\end {equation}
From (\ref{eq:terml}), we see that the
$m^2 \langle \hat G \v\cdot\E_\L \rangle$
term is $O(m^2 k^{-2} \D\cdot\E)$.
That's bigger than the $\D\cdot\E$ term on the left-hand side of
(\ref{eq:gauss2}) by a factor of $m^2/k^2$, and $m^2/k^2$
is large for the modes whose
physics I wish to correctly describe $(k \ll m)$.
So it is permissible, when implementing the constraints of
Gauss' Law, to ignore the contribution of the $\D\cdot\E$
on the left-hand side, leaving
\begin {equation}
   0 \simeq m^2 \langle \hat G (\v\cdot\E + \xi) \rangle .
\label{eq:gauss3}
\end {equation}
Rewriting back in terms of $W$, this is the $\omega \ll k$ equation
(\ref{eq:Rb}) presented earlier.

Finally, consider Ampere's Law (\ref{eq:Bc}).
For the moment, think about it in $A_0 = 0$ gauge, where it becomes
\begin {equation}
   \partial_t^2 \A + \D\times\D\times\A = m^2 \langle\v W \rangle .
\end {equation}
The first term is $O(\omega^2 A)$ and the second $O(k^2 A)$.
This suggests that one may drop the first term in comparison to the
second---at least in the transverse sector.
($\D\times\B = \D\times\D\times\A$ is
purely transverse).  The result is the equation (\ref{eq:Rc})
presented earlier.
For this equation to be consistent, it had better be that the
right-hand side is purely transverse as well (in the $\omega \ll k$ limit).
Indeed,
\begin {equation}
   \D\cdot\langle\v W\rangle
   = \langle \v\cdot\D W \rangle
   = \langle \v\cdot\E + \xi \rangle
   = \langle \xi \rangle ,
\label {eq:frog}
\end {equation}
where I've used the $\omega \ll k$ Boltzmann equation (\ref{eq:Ra}).
The $\v$-average $\langle \xi \rangle_\v$
of the noise $\xi$
vanishes for the following reason \cite{bodeker}.
Since $\xi$ is Gaussian noise, so is $\langle \xi \rangle_\v$.  But
\begin {equation}
   \Bigdlangle \langle \xi \rangle \, \langle \xi \rangle \Bigdrangle
   = \Bigl\langle \dlangle \xi(\v)\,\xi(\v') \drangle \Bigr\rangle_{\v,\v'}
   \propto \langle \deltaC(\v,\v') \rangle_{\v,\v'}
   = 0 ,
\end {equation}
so $\langle \xi \rangle$ is simply zero.
Then (\ref{eq:frog}) implies that $\langle \v W \rangle$
is indeed purely transverse in the $\omega \ll k$ effective theory.


\section{Two equations for $\A$}
\label{sec:twoeq}

There is a conceptual trap lurking in the $\omega \ll k$ equations
(\ref{eq:R}) that is easy to fall into.
Eq. (\ref{eq:Rb}) appears to say that $j^0 = m^2\langle W \rangle$
vanishes in the $\omega \to 0$ limit.  And so, by Gauss' Law,
that $\D\cdot\E$ = 0.  And one might take that to mean that longitudinal
electric fields $\E_\L$ are negligible compared to transverse fields $\E_\T$
in the $\omega \ll k$ limit.  This is incorrect.%
\footnote{
   Consider, for example, the case of $k \ll \gamma$, so that
   (\ref{eq:bodeker}) gives an effective description of the physics,
   but $k \gg g^2 T$, so that the physics is still perturbative.
   And consider, for example, the frequency scale
   $\omega \sim \sigma k^2$.  Then (\ref{eq:bodeker}) gives the
   order of magnitude relation $\sigma\E \sim \bzeta$, which means that
   all polarizations of $\E$ are the same order of magnitude.
   See ref.\ \cite{Blog2} for a detailed discussion of why the
   effective theory (\ref{eq:bodeker}) applies to the longitudinal as
   well as transverse sector.
}
The equation
(\ref{eq:Rb}) merely reflects the fact that the $\D\cdot\E$ term
in (\ref{eq:gauss2}) is negligible compared to the individual terms on
the right-hand side of that equation---there is no presumption about
how small $\E_\L$ is relative to $\E_\T$.  It is perhaps less confusing
to eliminate $W$ altogether, and replace eqs.\ (\ref{eq:R}) by
the two equations
\begin {mathletters}
\label{eq:noW0}
\begin {equation}
   0 = m^2 \langle \hat G(\v\cdot\E + \xi) \rangle ,
\end {equation}
\begin {equation}
   \D\times\B = m^2 \langle \v \hat G (\v\cdot\E + \xi) \rangle .
\end {equation}
\end {mathletters}%

The form (\ref{eq:R}) in terms of $W$ has the advantage of having a more direct
correspondence with the form of the original equations (\ref{eq:effB}).
I will want to refer to the $W$-eliminated form (\ref{eq:noW0}) in the
next section, however, and so it is useful to simplify the noise terms
in these equations.  In particular, the terms $m^2 \langle \hat G \xi \rangle$
and $m^2 \langle \v \hat G \xi \rangle$ are proportional to Gaussian noise
$\xi$ and so are themselves Gaussian noise, and Gaussian noise can be
completely specified just by specifying its correlator.
So, rewrite the two equations (\ref{eq:noW}) as
\begin {mathletters}
\label {eq:noW}
\begin {equation}
   0 = m^2 \langle \hat G \v \rangle \cdot\E + \eta ,
\label {eq:noWa}
\end {equation}
\begin {equation}
   \D\times\B = m^2 \langle \v \hat G \v \rangle \cdot\E + \bzeta_\T ,
\label{eq:noWb}
\end {equation}
\end {mathletters}%
where
\begin {mathletters}
\begin {equation}
   \dlangle \eta \eta \drangle
   = m^4 \langle \hat G \dlangle \xi \xi \drangle \hat G^\trans \rangle
   = 2 T m^2 \langle \hat G \, \deltaC \, \hat G^\trans \rangle ,
\end {equation}
\begin {equation}
   \dlangle \bzeta_\T \bzeta_\T \drangle
   = m^4 \langle \v \hat G \dlangle \xi \xi \drangle \hat G^\trans \v \rangle
   = 2 T m^2 \langle \v \hat G \, \deltaC \, \hat G^\trans \v \rangle .
\label{eq:zetaT1}
\end {equation}
\end {mathletters}%
The right-hand sides implicitly have factors of $\delta(t-t')$, which I
have suppressed.
The transpose on $\hat G$ indicates transposition in $\x$-space, color space,
and $\v$-space.  $D_i$ is the adjoint representation covariant derivative
and satisfies $D_i^\trans = - D_i$.  The linearized collision operator
$\deltaC$ is symmetric in $\v$-space since, by rotation invariance, it can
only depend%
\footnote{
   This argument assumes that collisions do not depend on spin, or that
   spin has been averaged over.  The $q \lesssim gT$ collisions that are of
   interest to this problem are indeed insensitive to spin at leading order
   in coupling.
}
on $\v\cdot\v'$.  [See (\ref{eq:deltaCleading}), for example, for the explicit
version at leading log order.]
So
\begin {equation}
   \hat G^\trans = (-\v\cdot\D + \deltaC)^{-1}
\end {equation}
Now note that
\begin {equation}
   \hat G \, \deltaC \, \hat G^\trans
   = \half \hat G \left[(\hat G^\trans)^{-1} + (\hat G)^{-1}\right]
                     \hat G^\trans
   = \half (\hat G + \hat G^\trans) ,
\end {equation}
so
\begin {equation}
   \dlangle \eta \eta \drangle
   = T m^2 \langle \hat G + \hat G^\trans \rangle ,
\end {equation}
\begin {equation}
   \dlangle \bzeta_\T \bzeta_\T \drangle
   = T m^2 \langle \v (\hat G + \hat G^\trans) \v \rangle .
\end {equation}
Taking $\v \to -\v$ in the $\v$ average shows that the $\hat G$ and
$\hat G^\trans$
terms give the same result, so that
\begin {mathletters}
\label{eq:noWnoise}
\begin {equation}
   \dlangle \eta \eta \drangle
   = 2T m^2 \, \langle \hat G \rangle .
\label{eq:eta}
\end {equation}
\begin {equation}
   \dlangle \bzeta_\T \bzeta_\T \drangle
   = 2T m^2 \, \langle \v \hat G \v \rangle .
\label{eq:zetaT}
\end {equation}
\end {mathletters}%
Finally, we can verify that $\eta$ and $\bzeta_\T$ are independent:
\begin {equation}
   \dlangle \eta \bzeta_\T \drangle
   = T m^2 \langle (\hat G + \hat G^\trans) \v \rangle
   = 0 ,
\end {equation}
where the last equality follows by $\v \to -\v$ in the term involving
$\hat G^\trans$.
The noise correlations (\ref{eq:noWnoise}), combined with the two
equations (\ref{eq:noW}) for $\A$, give a complete description of the
$\omega \ll k \ll m$ effective theory.


\section{One equation for $\A$}
\label{sec:combine}

It is possible to eliminate $W$ and write a single equation for
$A$ that embodies all of (\ref{eq:R}) or (\ref{eq:noW}).
Define the projection operator $\Po$ to be an operator in $\v$-space
that projects out functions that are independent of $\v$;
that is,
\begin {equation}
   \Po f(\v) = \langle f(\v) \rangle .
\end {equation}
The trick is to take the $1-\Po$ projection of (\ref{eq:Ra}), and
to note that $\Po$ vanishes on $\v\cdot\E$ and on $\xi$ (because
$\langle \xi \rangle = 0$, as explained earlier).  So
\begin {equation}
   (1-\Po) (\v\cdot\D + \deltaC) W = \v\cdot\E + \xi .
\label {eq:W1}
\end {equation}
Now note that (\ref{eq:Rb}) tells us that $\Po W = 0$, and so
\begin {equation}
   (1-\Po) (\v\cdot\D + \deltaC) (1-\Po) W = \v\cdot\E + \xi .
\end {equation}
Then we can solve for $W$ as
\begin {mathletters}
\label{eq:WG1}
\begin {equation}
   W = \hat G_1 (\v\cdot\E + \xi) ,
\end {equation}
\begin {equation}
   \hat G_1 \equiv \Bigl[(1-\Po) (\v\cdot\D + \deltaC) (1-\Po)\Bigr]^{-1} ,
\label {eq:G1a}
\end {equation}
where the inverse is understood to be taken in the space projected by
$1{-}\Po$.  An alternative way to obtain the same inverse is
\begin {equation}
   \hat G_1 = \lim_{\Lambda \to \infty}
                  (\v\cdot\D + \deltaC + \Lambda \Po)^{-1} .
\label {eq:G1b}
\end {equation}
\end {mathletters}%
Eq.\ (\ref{eq:WG1}) appears different from the solution (\ref{eq:W0}) used
earlier for $W$.  Indeed it {\it is}\/ different for arbitrary $\E$, but it
produces the
same $W$ when $\E$ is such that the $\omega \ll k$ Gauss' Law
(\ref{eq:Rb}) is satisfied.
The advantage of the present form is that it may be used to derive
a single equation containing all of the dynamics of the three equations
(\ref{eq:R}).
To proceed, use (\ref{eq:WG1}) in Ampere's Law (\ref{eq:Rc}) to get
(\ref{eq:Aeq}),
\begin {equation}
   \D \times \B = \sig(\D) \, \E + \bzeta ,
\label{eq:Aeq2a}
\end {equation}
where $\sig(\D)$ is a matrix in vector-index space,
\begin {equation}
   \sig_{ij}(\D) \equiv m^2 \langle v_i \hat G_1 v_j \rangle
       = \lim_{\Lambda \to \infty} m^2 \langle
               v_i (\v\cdot\D + \deltaC + \Lambda \Po)^{-1} v_j \rangle ,
\end {equation}
and $\bzeta$ is Gaussian noise given by
\begin {equation}
   \bzeta \equiv m^2 \langle \v \hat G_1 \xi \rangle
       = \lim_{\Lambda \to \infty} m^2 \langle
               \v (\v\cdot\D + \deltaC + \Lambda \Po)^{-1} \xi \rangle .
\label{eq:zeta}
\end {equation}

I'll show in a moment that (\ref{eq:Aeq2a}) subsumes the three equations
(\ref{eq:R}), but first I'll derive the correlation of the Gaussian noise
$\bzeta$.  Based on the analogy of (\ref{eq:Aeq2a}) with
the far-infrared effective theory (\ref{eq:bodeker}),
one might expect that the correlation is (\ref{eq:zz}).
To verify it, start from the definition
(\ref{eq:zeta}) of $\bzeta$, which gives
\begin {equation}
   \dlangle \zeta_i \zeta_j \drangle
   = m^4 \langle v_i \hat G_1 \dlangle \xi \xi \drangle \hat G_1^\trans v_j
        \rangle
   = 2T m^2 \, \langle v_i \hat G_1 \,\deltaC\, \hat G_1^\trans v_j
        \rangle .
\end {equation}
By arguments that parallel those used to derive the noise correlation
(\ref{eq:zetaT}) of $\bzeta_\T$ from the analogous starting point
(\ref{eq:zetaT1}) in the previous section, one obtains
\begin {equation}
   \dlangle \zeta_i \zeta_j \drangle
   = 2T m^2 \, \langle v_i \hat G_1 v_j \rangle
   = 2 T \, \sig(\D) .
\label {eq:zz2a}
\end {equation}

I'll now show that the simple equations I've derived,
\begin{mathletters}
\label{eq:combine2}
\begin {equation}
   \D \times \B = \sig(\D) \, \E + \bzeta ,
\label{eq:Aeq2}
\end {equation}
\begin {equation}
   \dlangle \bzeta \bzeta \drangle
   = 2T \, \sig(\D) ,
\label{eq:zz2}
\end {equation}
\end {mathletters}%
provide a complete description of the $\omega \ll k \ll m$
effective theory originally described by (\ref{eq:R}) and the correlation
(\ref{eq:xi}).  Specifically, I'll show how to recover the two individual
equations (\ref{eq:noW}) of the last section for Gauss' Law and Ampere's Law,
which were the result of trivially eliminating $W$ from (\ref{eq:R}).
It's convenient to first establish a relation between the
$(1-\Po)$ projected $W$ propagator $\hat G_1$ of (\ref{eq:G1a}) and
(\ref{eq:G1b}) and the original unprojected propagator of
(\ref{eq:G}).  The relation is
\begin {equation}
   \hat G_1 = \hat G - \hat G \Po \langle \hat G \rangle^{-1} \hat G ,
\label {eq:G1toG}
\end {equation}
which can also be thought as of the rule
\begin {equation}
   \hat G_1 - \hat G =
        - \hat G \rangle \, \langle \hat G \rangle^{-1} \,
                \langle \hat G
\end {equation}
for the purpose of lexical substitution in formulas,
{\it e.g.}
\begin {equation}
   \sig(\D) = m^2 \langle \v \hat G_1 \v \rangle
   = m^2 \left[ \langle \v \hat G \v \rangle
          - \langle \v \hat G \rangle \langle \hat G \rangle^{-1}
                 \langle \hat G \v \rangle \right] .
\label{eq:sig2}
\end {equation}
It's easy to verify (\ref{eq:G1toG}) by first checking that it lives
in the space projected by $1-\Po$,
\begin {equation}
   \Po \left[\hat G - \hat G \Po \langle G \rangle^{-1} \hat G\right]
   = \Po G - \Po \langle G \rangle \langle G \rangle^{-1} \hat G
   = \Po G - \Po G
   = 0
\end {equation}
(and similarly
$\left[\hat G + \hat G \Po \langle G \rangle^{-1} \hat G\right]
\Po=0$),
and then checking explicitly that it's the desired inverse:
\begin {eqnarray}
   \left[\hat G - \hat G \Po \langle G \rangle^{-1} \hat G\right]
      \hat G_1^{-1}
   &=& (1-\Po)
        \left[\hat G - \hat G \Po \langle G \rangle^{-1} \hat G\right]
        \hat G^{-1} (1-\Po)
\nonumber\\
   &=& (1-\Po) \left[1 - \hat G \Po \langle G \rangle^{-1}\right]
        (1-\Po)
\nonumber\\
   &=& (1-\Po) .
\end {eqnarray}

Now I'll show that the gauge field Langevin equation (\ref{eq:Aeq2}) implies
Gauss' Law (\ref{eq:noWa}) by dotting $\langle G \rangle \D$ into both sides
of (\ref{eq:Aeq2}):
\begin {equation}
   0 = \langle G \rangle \left[ \D \,\sig(\D)\, \E + \D \cdot \bzeta \right] .
\label{eq:getgauss1}
\end {equation}
Using (\ref{eq:sig2}),
\begin {equation}
   \langle G \rangle \D \,\sig(\D)\, \E
   = m^2 \langle G \rangle
    \left[ \langle \v \cdot \D \hat G \v \rangle
        - \langle \v\cdot\D \hat G \rangle \langle \hat G\rangle^{-1}
              \langle \hat G \v \rangle
     \right] \E .
\label {eq:getgass2}
\end {equation}
We can simplify using the trick discussed earlier,
\begin {equation}
   \langle \v \cdot\D \hat G
   = \langle (\v\cdot\D + \deltaC) \hat G
   = \langle \, ,
\label{eq:trickG1}
\end {equation}
so that
\begin {mathletters}
\label {eq:trickG2}
\begin {equation}
   \langle \v\cdot\D \hat G \v \rangle = \langle \v \rangle = 0 ,
\end {equation}
\begin {equation}
   \langle \v\cdot\D \hat G \rangle = 1 .
\end {equation}
\end {mathletters}%
Eq.\ (\ref{eq:getgass2}) for the $\sig$ term then becomes
\begin {equation}
   \langle G \rangle \D \,\sig(\D)\, \E
   = - m^2 \langle \hat G \v \rangle \cdot \E .
\end {equation}
So (\ref{eq:getgauss1}) becomes
\begin {equation}
   0 = m^2 \langle \hat G \v \rangle \cdot\E
         - \langle G \rangle \D \cdot \bzeta .
\label{eq:getgauss3}
\end {equation}
Compare to Gauss' Law (\ref{eq:noWa}).
The last term is Gaussian noise, and all that matters for the purpose
of reproducing Gauss' Law (\ref{eq:noWa}) is to check that
$\eta' \equiv - \langle G \rangle \D \cdot\bzeta$ has the same noise
correlation (\ref{eq:eta}) as $\eta$ of the last section.
First put
together (\ref{eq:G1toG}) and (\ref{eq:trickG1}) to get
\begin {equation}
   \langle \v\cdot\D \hat G_1
   = \Bigl\langle \left[ 1 - \langle \hat G \rangle^{-1} \hat G \right] .
\end {equation}
Then, using the $\bzeta$ correlation (\ref{eq:zz2}),
\begin {eqnarray}
   \dlangle \eta' \eta' \drangle
   &=& \langle \hat G \rangle \D \cdot \dlangle \bzeta \bzeta \drangle
            \cdot \D^\trans \langle \hat G^\trans \rangle
\nonumber\\
   &=& - 2 T m^2 \langle \hat G \rangle
        \langle \v\cdot\D \hat G_1 \v\cdot\D \rangle
        \langle \hat G \rangle
\nonumber\\
   &=& 2 T m^2 \langle \hat G \v\cdot\D \rangle \langle \hat G \rangle
\nonumber\\
   &=& 2 T m^2 \langle \hat G \rangle
   = \dlangle \eta \eta \drangle .
\end {eqnarray}
Alternatively, one could go back to the expression (\ref{eq:zeta})
for $\bzeta$ in terms of $\xi$, and show directly that
$\eta' \equiv - \langle G \rangle \D \cdot\bzeta$ is the same
as the $\eta \equiv m^2 \langle \hat G \xi \rangle$.

Having obtained Gauss' Law, we can now check that the single equation
(\ref{eq:Aeq2}) also enforces Ampere's Law (\ref{eq:noWb}).
Expand $\sig$ using (\ref{eq:sig2}), so that (\ref{eq:Aeq2}) becomes
\begin {equation}
   \D\times\B
   = m^2 \langle \v \hat G \v \rangle \cdot \E
                - m^2 \langle \v \hat G \rangle
                      \langle \hat G \rangle^{-1} \langle \hat G \v \rangle
                      \cdot\E
                + \bzeta
   = m^2 \langle \v \hat G \v \rangle \cdot \E + \bzeta'_\T ,
\end {equation}
where the last equality uses Gauss' Law (\ref{eq:getgauss3}) and defines%
\begin {equation}
   \bzeta'_\T \equiv \bzeta
   - \langle \v G \rangle \D\cdot\bzeta .
\end {equation}
One may verify that the noise $\bzeta'_\T$ is equivalent to the noise
$\bzeta_\T$ (\ref{eq:zetaT}) of the previous section.
It is worth mentioning that $\bzeta'_\T$ is transverse
($\D\cdot\bzeta'_\T = 0$), but it is {\it not}\/
simply the transverse projection $\PT \bzeta$ of $\bzeta$.


\section {Path integrals and ambiguities}
\label{sec:path}

\subsection {A warm-up: the \boldmath$k \ll \gamma$ theory}

\subsubsection {The path integral in $A_0=0$ gauge}

To warm up to talking about path integral formulations for the
$\omega\ll k\ll m$ Langevin equation (\ref{eq:Aeq}), I will start by
discussing path integrals for the simpler, more infrared effective theory
described by (\ref{eq:bodeker}) for $\omega \ll k \ll \gamma$.
In $A_0 = 0$ gauge,
\begin {mathletters}
\label{eq:bodeker2}
\begin {equation}
   \sigma \dot\A = - \D\times\B + \bzeta ,
\end {equation}
\begin {equation}
    \dlangle \bzeta \bzeta \drangle = 2 \sigma T .
\end {equation}
\end {mathletters}%
In this gauge, the above Langevin equation has a nice physical
interpretation, because it can be rewritten as
\begin {equation}
   \sigma \dot\A = - {\delta\over\delta \A} \, {\cal V[\A]}
                   + \bzeta ,
\end {equation}
where
\begin {equation}
   {\cal V}[\A] = \int d^3x \> \half \B^a \cdot \B^a
\end {equation}
is the magnetic energy.  This means the Langevin equation is just an
infinite degree of freedom version of
the kinematics of a highly damped particle in a
potential $V(\q)$:
\begin {mathletters}
\label{eq:finite}
\begin {equation}
   \sigma \dot q_i = - {d\over d q_i} \, V(\q) + \zeta_i ,
\label {eq:finitea}
\end {equation}
\begin {equation}
   \dlangle \zeta_i(t) \,  \zeta_j(t') \drangle
   = 2 \sigma T \delta_{ij} \, \delta(t-t') .
\end {equation}
\end {mathletters}%
It is well known how to rewrite such equations, and their field theory
counterparts, as path integrals,%
\footnote{
   For a review, see, for example, chapters 4 and 17 of ref.\
   \cite{ZinnJustin}.
}
but I'll briefly review the steps here.
Keep to the notation (\ref{eq:finite}) for the moment, and
first consider an integral over the distribution for the Gaussian noise:
\begin {equation}
   Z \equiv \int [{\cal D}\bzeta(t)] \>
         \exp\!\left[ - {1\over 4 \sigma T} \int dt\> |\bzeta(t)|^2 \right] .
\end {equation}
Now insert a factor of one in the form of the equation of motion
(\ref{eq:finitea}):
\begin {equation}
   Z = \int [{\cal D}\bzeta(t)] \>
         \exp\left[ - {1\over 4 \sigma T} \int dt\> |\bzeta(t)|^2 \right] 
         \> \int [{\cal D}\q(t)] \>
         \delta\!\left[ \sigma \dot \q + \grad_\q \, V(\q) - \bzeta \right]
         J[\q] ,
\label {eq:patha}
\end {equation}
where the $\delta$ function is functional, and the corresponding
Jacobian is given by a functional determinant
\begin {eqnarray}
   J[\q] &\equiv& \det_{ij} \left( {d \over d q_i}
               \left[ \sigma \dot q_j + \nabla_{q_j} \, V(\q) - \bzeta \right]
               \right)
\nonumber\\
   &=& \det_{ij} \left( \sigma \delta_{ij} {d\over dt}
              + \nabla_{q_i} \nabla_{q_j} \, V(\q) \right)
\nonumber\\
   &=& \int [{\cal D}\bar\c] \> [{\cal D}\c] \>
         \exp\left[ - \int dt \> \bar b_i \left( \sigma \delta_{ij} \partial_t
              + \nabla_{q_i} \nabla_{q_j} \, V(\q) \right) b_j \right] .
\end {eqnarray}
I have introduced ghosts $\bar\b$ and $\b$ in the last line, which
are not related to gauge fixing.
Now use the $\delta$-function to perform the noise
integral in (\ref{eq:patha}), to get
\begin {equation}
   Z = \int [{\cal D}\q(t)] \>
         J[\q] \,
         \exp\left[ - {1\over 4 \sigma T} \int dt\>
                \left|\sigma\dot\q + \grad_\q V(\q) \right|^2 \right] ,
\label {eq:pathb}
\end {equation}

One may simplify the Jacobian further, but the details depend on how one
regularizes short times in the path integral.  That is, there is
sensitivity
to what convention one uses for discretizing time in the path integral.
[In contrast, the original Langevin equation (\ref{eq:finite}) is
insensitive to the details of short-time regularization.]
If one makes the standard choice of a time-symmetric discretization
scheme, where $\dot\q$ and $\q$ in the path integral are interpreted as
\begin {equation}
   \dot\q = {q(t_i) - q(t_{i-1}) \over \Delta t} ,
   \qquad
   \q = {q(t_i) + q(t_{i-1}) \over 2} ,
\label {eq:symdis}
\end {equation}
then one may show that the Jacobian simplifies to
\cite{ZinnJustin,theta zero}
\begin {equation}
   J[\q] = \exp\left[ - {\theta(0)\over \sigma} \int dt \>
               \nabla_\q^2 V(\q) \right]
\end {equation}
with the symmetric interpretation
\begin {equation}
   \theta(0) = \half
\end {equation}
of the step function $\theta(t)$.

In field theory, $\q$ becomes $\A$, and $\q$-derivatives become functional
derivatives.  The path integral (\ref{eq:pathb}) becomes, in the case at
hand,
\begin {equation}
   Z = \int [{\cal D}\A] \>
         J[\A] \,
         \exp\left[ - {1\over 4 \sigma T} \int dt\> d^3x\>
                \left|\sigma\dot\A + \D\times\B \right|^2 \right] ,
\label {eq:pathA0}
\end {equation}
\begin {eqnarray}
   J[\A] &=& \exp\left[ - {\theta(0)\over \sigma} \int dt \> d^3x \>
               {\delta\over\delta A_i^a(\x)}\, (\D\times\B)_i^a(\x) \right]
\nonumber\\
   &=& \exp\left[ - {\theta(0)\over \sigma} 
               \, \delta^{(3)}(0) \int dt \> d^3x \>
               {d\over dA_i^a}\, (\D\times\B)_i^a \right] .
\end {eqnarray}
It's easy enough to take the derivative, to get
\begin {equation}
   J[\A] = \exp\left[ - \sigma^{-1} \delta^{(3)}(0) \, \tr \D^2 \right] ,
\label {eq:J}
\end {equation}
but it is unnecessary if one uses dimensional regularization.  In dimensional
regularization, $\delta^{(d)}(0)$ vanishes (where I take $d = 3-\eps$
to be the number of spatial dimensions), and so
\begin {equation}
   J[\A] = 1 \qquad \mbox{(dimensional regularization)} .
\end {equation}
This is a feature of {\it any}\/ Langevin field
equation that is local in space.


\subsubsection {The path integral in other gauges}

Knowing the result in $A_0=0$ gauge, it is easy to guess the
corresponding path integral {\it without}\/ gauge-fixing:%
\footnote{
   See the discussion surrounding eq.\ (4.9) of
   ref.\ \cite{flow gauges}.
}
\begin {equation}
   Z = \int [{\cal D} A_0] \> [{\cal D}\A] \>
         J[\A] \,
         \exp\left[ - {1\over 4 \sigma T} \int dt\> d^3x\>
                \left|\sigma\E + \D\times\B \right|^2 \right] .
\end {equation}
This can be verified by now fixing $A_0=0$ gauge in the usual way,
and obtaining (\ref{eq:pathA0}).
The advantage of the gauge-invariant form is that one can now alternatively fix
other gauges in the usual way, by introducing Faddeev-Popov ghosts
$\c$ and $\bar\c$.  For example, to fix Coulomb gauge,
\begin {equation}
   Z = \int [{\cal D}A_0] [{\cal D}\A]
          [{\cal D}\bar\c] [{\cal D} \c]
          \> \delta(\grad\cdot\A) \> J[\A] \>
          \exp\left(-\int dt \> d^3 x \> L_{\rm Coloumb}\right) ,
\end {equation}
\begin {equation}
   L_{\rm Coloumb} = {1\over 4\sigma T} \left[
          \left| -\sigma\E + \D\times\B \right|^2
          + \bar\c \grad\cdot\D \c
   \right] .
\label {eq:Lcoulomb}
\end {equation}


\subsection {The \boldmath$k \ll m$ theory}

There are two important differences between the $k \ll m$ Langevin equation
(\ref{eq:combine}) and the simpler $k \ll \gamma$ equation
(\ref{eq:bodeker}).
The first is that the $k \ll m$ equation is non-local, which means that
the Jacobian term in the path integral, analogous to (\ref{eq:J}), will not
involve $\delta^{(3)}(0)$ and so will not trivially vanish in dimensional
regularization.
The second is that the amplitude of the damping and the noise in
the $k \ll m$ equation depends on the state $\A$ of the system.
As mentioned before, this means that the continuum Langevin equation
does not have a well-defined meaning.
The Langevin equation itself (and not just the path integral description)
is sensitive to the ultraviolet and
details of UV frequency regularization.

The fact that an effective theory is sensitive to details of ultraviolet
regularization is not novel.  Almost all effective field theories require
ultraviolet regularization, and it was only the anomalous fact that the
$k \ll \gamma$ effective theory (\ref{eq:bodeker}) happens to be
ultraviolet {\it finite}%
\footnote{
  For a discussion in the present context, see ref.\ \cite{Blog1}.
}
that meant we didn't need to regularize it.  As usual, one should
simply pick a regularization scheme and then fix the regularized
parameters of the effective theory so that it reproduces the infrared
physics of whatever more fundamental theory underlies it.
In ref.\ \cite{langevin}, I have discussed this matching problem for general
systems of the form
\begin {mathletters}
\label{eq:multnoise}
\begin {equation}
   \sigma_{ij}(\q) \, \dot q_j = - \nabla_{q_i} V(\q) + \zeta_i ,
\end {equation}
\begin {equation}
   \dlangle \zeta_i(t) \, \zeta_j(t') \drangle
        = 2 T \, \sigma_{ij}(\q) \, \delta(t-t') ,
\end {equation}
\end {mathletters}%
in cases where it is known that the equilibrium distribution for $\q$ is
\begin {equation}
   P_{\rm eq}(\q) = e^{-V(\q)/T}
\label{eq:Peq}
\end {equation}
in whatever approximation one is working in.
This is useful in the present case because static equilibrium properties
of hot gauge theories are much simpler to analyze than dynamical ones,
and indeed the equilibrium
distribution in $A_0=0$ gauge should be (\ref{eq:Peq}) with $V$ the
magnetic energy.
In ref.\ \cite{langevin}, I discuss how knowledge of the equilibrium
distribution (\ref{eq:Peq}) forces the ambiguities inherent in the
continuum Langevin equations (\ref{eq:multnoise}) to be resolved in a
particular way.  I also showed
that the corresponding path integral formulation is
\begin {equation}
   Z = \int [{\cal D}\q(t)] \exp\left[-\int dt \> L(\dot\q,\q) \right] ,
\end {equation}
\begin {equation}
  L(\dot\q, \q) = {1\over 4T} \, (\g \dot \q + \grad_{\q} V)^\trans \g^{-1}
                                 (\g \dot \q + \grad_{\q} V)
      + L_1(\q) ,
\label {eq:L}
\end {equation}
\begin {equation}
  L_1(\q) = 
      - {1\over2} \nabla_{q_i} [(\g^{-1})_{ij} \nabla_{q_j} V]
      + {T\over4} \nabla_{q_i} \nabla_{q_j} (\g^{-1})_{ij}
      - {1\over2} \, \delta(0) \, \tr \, \ln \g \,,
\label {eq:L1}
\end {equation}
if the path integral is defined with symmetric time discretization
(\ref{eq:symdis}).  $\delta(0)$ above is short-hand for
$\delta(t{=}0) = (\Delta t)^{-1}$.
The first term in (\ref{eq:L}) is the obvious generalization of
the exponent in (\ref{eq:pathb}) from scalar $\sigma$
to matrix $\sigma_{ij}(\q)$.
The remaining $L_1(\q)$ term represents the appropriate Jacobian
(more accurately, $-\ln J$) and the terms necessary for the
desired resolution of the
ambiguities of the continuum Langevin equation (\ref{eq:multnoise}).
One may easily verify that specialization to the case
$\sigma_{ij}(\q) = \sigma \delta_{ij}$, with $\sigma$ constant,
reproduces the earlier result (\ref{eq:pathb})
[up to an irrelevant constant normalization].

In $A_0=0$ gauge, we can now obtain the path integral for the gauge theory
case by replacing
$\q$ by $\A$ and derivatives by functional derivatives.  The resulting
action density is
\begin {equation}
   L = {1\over 4T}
           \Bigl[ \sig(\D)\, \dot\A + \D\times\B \Bigr]^{\rm T}
           \sig(\D)^{-1}
           \Bigl[ \sig(\D)\, \dot\A + \D\times\B \Bigr]
    + L_1[\A],
\label{eq:LA0}
\end {equation}
which, except for the $L_1[\A]$ term,
is the natural generalization of the $k\ll \gamma$ action
in (\ref{eq:pathA0}).
The $L_1[\A]$ term, however, is ugly as sin.
So much so, that it is unilluminating to write it down, other than
to refer back to the discrete version (\ref{eq:L1}).
I have been unable to find an attractive form for $L_1[\A]$.
Fortunately, $L_1[\A]$ does not enter at all into certain important
applications of this formalism, as I will discuss shortly.

A gauge-unfixed version of the action (\ref{eq:L}) can be found simply
by finding a gauge-invariant action that becomes (\ref{eq:L}) when
fixed to $A_0=0$ gauge.  The result is
\begin {equation}
   L = {1\over 4T}
           \Bigl[ -\sig(\D)\, \E + \D\times\B \Bigr]^{\rm T}
           \sig(\D)^{-1}
           \Bigl[ -\sig(\D)\, \E + \D\times\B \Bigr]
    + L_1[\A] .
\label {eq:Lfinal}
\end {equation}
This action may then be used to fix whatever gauge is desired.

Note that $L_1[\A]$, though derived in $A_0=0$ gauge, is guage-invariant
under general time-dependent gauge transformations.
The derivation in $A_0=0$ gauge implied that $L_1[\A]$ is invariant under
time-{\it independent} gauge transformations.
Because $L_1[\A]$ does not involve any time derivatives (and because I did
not introduce $A_0$ into this term), it is then
automatically invariant under time-dependent transformations as well.


\subsection {The nature of \boldmath$L_1[\A]$}

The coupling constant $g$ is a convenient parameter for counting powers of
the loop expansion.  At high temperature, the parameter which controls
the effectiveness of the loop expansion is not $g^2$ by itself, but it is
at least proportional to an explicit factor of $g^2$.
For analysis of static equilibrium quantities, for example, the loop
expansion parameter is $g^2 T / k$ (once appropriate resummations have
been implemented) for momenta $k \gtrsim g^2 T$.
The fact that physics is somehow treatable perturbatively
for $k \gg g^2T$ (after integrating out degrees of freedom that decouple
at various physical thresholds) is a reflection of the fact that the size of
gauge field fluctuations is perturbatively small for such $k$.
In refs.\ \cite{overview,sigma}, for example, Yaffe and I use the
loop expansion of the $\omega \ll k \ll m$ theory at $k \sim \gamma$
to compute corrections to color conductivity and hot electroweak baryon
number violation.  The loop expansion is in that case an expansion in
$g^2 T / k \sim [\ln(1/g)]^{-1}$.

To understand at what order in the loop expansion interactions in the
Lagrangian might contribute, it is therefore important to understand
what explicit factors of $g$ are associated with those interactions.
Let's focus in particular on the (horrible) terms of $L_1[\A]$.
First, note that $\A$ only appears in the combination $g\A$ in
$\sigma(\D)$, since $\D = \grad + g\A$.  So expanding $\sigma(\D)$ in powers
of $g$ gives
\begin {equation}
   \sigma(\D) = \sigma(\grad) + O(g A) + O(g^2 A^2) + \cdots,
\end {equation}
and then similarly,
\begin {equation}
   [\sigma(\D)]^{-1} = [\sigma(\grad)]^{-1} + O(g A) + O(g^2 A^2) + \cdots,
\end {equation}
where I am only keeping track of the explicit powers of
$g$ and $\A$ at each order.
The terms in this expansion are {\it not}\/ local in space.

One can now read off, for instance, that the gauge theory term corresponding to
the $\nabla_{q_i} \nabla_{q_j} (\sigma^{-1})_{ij}$ term in (\ref{eq:L1}) for
$L_1$ must be schematically of the form
\begin {equation}
   {\delta^2 \over \delta A^2} \sigma^{-1} = O(g^2) + O(g^3 A) + O(g^4 A^2)
              + \cdots .
\end {equation}
The $O(g^2)$ term is independent of $\A$ and so can be discarded.
There can't actually be an $O(g^2 A)$ term in the Lagrangian because
there's no way for it to be a color singlet.  So the leading
piece of the
$\delta^2 \sigma^{-1} / \delta A^2$ term of $L_1[\A]$ is $O(g^4 A^2)$.
The explicit factor of $g^4$ implies that it will be
suppressed by two powers of the loop expansion
parameter compared to the $O(g^0 A^2)$ terms in the action
(\ref{eq:Lfinal}), which determine the $\A$ propagator.
[And the $\A^3$ and so forth terms are similarly suppressed compared to
non-$L_1$ $\A^3$ and so forth terms in (\ref{eq:Lfinal}).]

Similarly, the magnetic energy is
\begin {equation}
   V = O(A^2) + O(g A^3) + O(g^2 A^4) .
\end {equation}
The possible terms arising from the $\nabla_\q [\sigma^{-1} \nabla_\q V]$
term in (\ref{eq:L1}) are then of the form
\begin {equation}
   {\delta\over\delta A} \left[ \sigma^{-1} {\delta\over\delta A} V \right]
   = O(1) + O(g A) + O(g^2 A^2) + \cdots .
\end {equation}
Again, $O(1)$ can be discarded, and $O(g A)$ can't appear in the action,
so the leading term in $g$ must be $O(g^2 A^2)$.
This is suppressed by one power
of the loop expansion parameter compared to the $O(g^0 A^2)$ terms in the
action (\ref{eq:Lfinal}).
The analysis of the remaining term, $\ln\sigma$, is similar.

The conclusion is that the interactions among $\A$ generated by
$L_1[\A]$ will all be suppressed by at least one power of the loop
expansion parameter, compared to those appearing in the other terms
of (\ref{eq:Lfinal}).
In ref.\ \cite{sigma}, Yaffe and I show that this suppression is
enough to permit a next-to-leading-log order analysis of the color
conductivity and the hot electroweak baryon number violation rate
without requiring use of an explicit form for $L_1[\A]$.

A word of caution about the above analysis is required, however.
The correspondence between explicit powers of $g^2$ and the expansion
parameter $g^2 T / k$ only works if one has an effective theory that
properly integrates out all of the physics above the scale you are
interested in.  For example, if one does perturbation theory in the
original $k \ll T$ hard-thermal loop effective theory
(\ref{eq:htl}), the loop expansion will break down at $k \sim \gamma$:
the loop expansion parameter will be $O(1)$ instead of
order $g^2T/k \sim g^2T/\gamma \sim [\ln(1/g)]^{-1}$.
One must instead make the loop expansion in the $k \ll m$ effective theory,
which incorporates the effects of collisions into the bare
propagators and vertices.
As long as the correct effective theory is used, there {\it should}\/
be no problem.  However, for the sake of caution, it would be useful
to have a much more explicit analysis of the suppression of the
$L_1[\A]$ terms than I have been able to give.



\begin {references}

\bibitem {bodeker}
    D. B\"odeker,
    {\tt hep-ph/9801430},
    {\sl Phys.\ Lett.}\ {\bf B426}, 351 (1998);
    {\tt hep-ph/9905239}.

\bibitem{braaten&pisarski}
    E. Braaten and R. Pisarski,
    Phys.\ Rev.\ {\bf D45}, 1827 (1992);
    Nucl.\ Phys. {\bf B337}, 569 (1990).

\bibitem{adjXadj}
   S. Mr\'{o}wczy\'{n}ski,
   {\sl Phys.\ Rev.}\ {\bf D39}, 1940 (1989);
   H.-Th. Elze and U. Heinz,
   {\sl Phys.\ Rept.}\ {\bf 183}, 81 (1989);
   J. Blaizot and E. Iancu,
   {\sl Nucl.\ Phys.}\ {\bf B417}, 609 (1994);
   and references therein.

\bibitem {W}
   J. Blaizot and I. Iancu,
   {\sl Phys.\ Rev.\ Lett.}\ {\bf 72}, 3317 (1994).

\bibitem {asy}
  P. Arnold, D. Son, and L. Yaffe,
  {\tt hep-ph/9609481},
  Phys.\ Rev.\ {\bf D55}, 6264 (1997).

\bibitem{pi}
   H. Weldon,
     Phys.\ Rev.\ {\bf D26}, 1394 (1982);
   U. Heinz,
     Ann.\ Phys.\ (N.Y.) {\bf 161}, 48 (1985); {\bf 168}, 148 (1986).

\bibitem {Blog1}
   P. Arnold, D. Son, and L. Yaffe,
   {\tt hep-ph/9810216},
   Phys.\ Rev.\ {\bf D59}, 105020 (1999);

\bibitem{manuelFD}
   D. Litim and C. Manuel, {\tt hep-ph/9910348}.

\bibitem{mooreBlog}
   G. Moore,
   {\tt hep-ph/9810313}.

\bibitem{overview}
   P. Arnold and L. Yaffe,
      {\it Non-perturbative dynamics of hot non-Abelian gauge fields:
      Beyond leading log},
      {\tt hep-ph/9912305}.

\bibitem {sigma}
   P. Arnold and L. Yaffe,
      {\it High temperature color conductivity at next-to-leading log order},
      {\tt hep-ph/9912306}.

\bibitem {Blog2}
   P. Arnold, D. Son, and L. Yaffe,
   {\tt hep-ph/9901304},
   Phys.\ Rev.\ {\bf D60}, 025007 (1999).

\bibitem {ZinnJustin}
    J. Zinn-Justin, {\sl Quantum Field Theory and Critical Phenomena},
    2nd edition (Oxford University Press, 1993).

\bibitem{theta zero}
    F. Langouche, D. Roekaerts, and E. Tirapegui,
      Physics {\bf 95A}, 252 (1979);
    H. Kawara, M. Namiki, H. Okamoto, and S. Tanaka,
      Prog.\ Theor.\ Phys.\ {\bf 84}, 749 (1990);
    N. Komoike,
      Prog.\ Theor.\ Phys.\ {\bf 86}, 575 (1991).

\bibitem {flow gauges}
   H. Chan and M. Halpern,
   Phys.\ Rev.\ {\bf D33}, 540 (1985).

\bibitem {langevin}
   P. Arnold,
      {\it Langevin equations with multiplicative noise: resolution of
      time discretization ambiguities for equilibrium systems},
      hep-ph/9912208.

\end {references}
\end {document}